\documentclass[apj]{emulateapj}
\usepackage{graphicx}
\usepackage{bm}              
\usepackage{graphics}
\usepackage{epsfig}
\usepackage{ulem}
\usepackage{symb}
\usepackage{dcolumn}
\usepackage{bm}
\usepackage{epsf}
\usepackage{hyperref}
\usepackage{wrapfig}
\usepackage{hyperref,ifthen}
\usepackage{rotating,makecell} 
\newcommand{\adsurl}[1]{\href{#1}{ADS}}
\providecommand{\url}[1]{\href{#1}{#1}}

\large\newlength{\oldparskip}\parskip=.1in
\newcommand{\be}{\begin{equation}}
\newcommand{\ee}{\end{equation}}
\newcommand{\ba}{\begin{eqnarray}}
\newcommand{\ea}{\end{eqnarray}}

\newcommand{\mpc}{Mpc$^{-1}$}
\newcommand{\wmap}    {{WMAP}}
\newcommand{\act}    {ACT}

\newcommand{\arone}    {148 GHz}
\newcommand{\artwo}  {218 GHz}

\begin{document}

\title{The Atacama Cosmology Telescope: a measurement of the primordial power spectrum}

\author{
Ren\'ee~Hlozek\altaffilmark{1},
Joanna~Dunkley\altaffilmark{1,2,3},
Graeme~Addison\altaffilmark{1},
John~William~Appel\altaffilmark{2},
J.~Richard~Bond\altaffilmark{4},
C.~Sofia~Carvalho\altaffilmark{5},
Sudeep~Das\altaffilmark{6,2,3},
Mark~J.~Devlin\altaffilmark{7},
Rolando~D\"{u}nner\altaffilmark{8},
Thomas~Essinger-Hileman\altaffilmark{2},
Joseph~W.~Fowler\altaffilmark{2,9},
Patricio~Gallardo\altaffilmark{8},
Amir~Hajian\altaffilmark{4,3,2},
Mark~Halpern\altaffilmark{10},
Matthew~Hasselfield\altaffilmark{10},
Matt~Hilton\altaffilmark{11},
Adam~D.~Hincks\altaffilmark{2},
John~P.~Hughes\altaffilmark{12},
Kent~D.~Irwin\altaffilmark{9},
Jeff~Klein\altaffilmark{7},
Arthur~Kosowsky\altaffilmark{13},
Tobias~A.~Marriage\altaffilmark{14,3},
Danica~Marsden\altaffilmark{7},
Felipe~Menanteau\altaffilmark{12},
Kavilan~Moodley\altaffilmark{15},
Michael~D.~Niemack\altaffilmark{9,2},
Michael~R.~Nolta\altaffilmark{4},
Lyman~A.~Page\altaffilmark{2},
Lucas~Parker\altaffilmark{2},
Bruce~Partridge\altaffilmark{16},
Felipe~Rojas\altaffilmark{8},
Neelima~Sehgal\altaffilmark{17},
Blake~Sherwin\altaffilmark{2},
Jon~Sievers\altaffilmark{4},
David~N.~Spergel\altaffilmark{3},
Suzanne~T.~Staggs\altaffilmark{2},
Daniel~S.~Swetz\altaffilmark{7,9},
Eric~R.~Switzer\altaffilmark{18,2},
Robert~Thornton\altaffilmark{7,19},
Ed~Wollack\altaffilmark{20}}
\altaffiltext{1}{Department of Astrophysics, Oxford University, Oxford, 
UK OX1 3RH}
\altaffiltext{2}{Joseph Henry Laboratories of Physics, Jadwin Hall,
Princeton University, Princeton, NJ, USA 08544}
\altaffiltext{3}{Department of Astrophysical Sciences, Peyton Hall, 
Princeton University, Princeton, NJ USA 08544}
\altaffiltext{4}{Canadian Institute for Theoretical Astrophysics, University of
Toronto, Toronto, ON, Canada M5S 3H8}
\altaffiltext{5}{ IPFN, IST, Av. RoviscoPais, 1049-001Lisboa, Portugal \& RCAAM, Academy of Athens, Soranou Efessiou 4, 11-527 Athens, Greece}
\altaffiltext{6}{Berkeley Center for Cosmological Physics, LBL and
Department of Physics, University of California, Berkeley, CA, USA 94720}
\altaffiltext{7}{Department of Physics and Astronomy, University of
Pennsylvania, 209 South 33rd Street, Philadelphia, PA, USA 19104}
\altaffiltext{8}{Departamento de Astronom{\'{i}}a y Astrof{\'{i}}sica, 
Facultad de F{\'{i}}sica, Pontific\'{i}a Universidad Cat\'{o}lica de Chile,
Casilla 306, Santiago 22, Chile}
\altaffiltext{9}{NIST Quantum Devices Group, 325
Broadway Mailcode 817.03, Boulder, CO, USA 80305}
\altaffiltext{10}{Department of Physics and Astronomy, University of
British Columbia, Vancouver, BC, Canada V6T 1Z4}
\altaffiltext{11}{School of Physics and Astronomy, University of Nottingham, University Park, Nottingham, NG7 2RD}
\altaffiltext{12}{Department of Physics and Astronomy, Rutgers, 
The State University of New Jersey, Piscataway, NJ USA 08854-8019}
\altaffiltext{13}{Department of Physics and Astronomy, University of Pittsburgh, 
Pittsburgh, PA, USA 15260}
\altaffiltext{14}{Dept. of Physics and Astronomy, The Johns Hopkins University, 3400 N. Charles St., Baltimore, MD 21218-2686}
\altaffiltext{15}{Astrophysics and Cosmology Research Unit, School of
Mathematical Sciences, University of KwaZulu-Natal, Durban, 4041,
South Africa}
\altaffiltext{16}{Department of Physics and Astronomy, Haverford College, Haverford, PA, USA 19041}
\altaffiltext{17}{Kavli Institute for Particle Astrophysics and Cosmology, Stanford
University, Stanford, CA, USA 94305-4085}
\altaffiltext{18}{Kavli Institute for Cosmological Physics, 
Laboratory for Astrophysics and Space Research, 5620 South Ellis Ave.,
Chicago, IL, USA 60637}
\altaffiltext{19}{Department of Physics , West Chester University 
of Pennsylvania, West Chester, PA, USA 19383}
\altaffiltext{20}{Code 553/665, NASA/Goddard Space Flight Center,
Greenbelt, MD, USA 20771}

\begin{abstract}
We present constraints on the primordial power spectrum of adiabatic fluctuations using data from the 2008 Southern Survey of the Atacama Cosmology Telescope (ACT). The angular resolution of ACT provides sensitivity to scales beyond $\ell = 1000$ for resolution of multiple peaks in the primordial temperature power spectrum, which enables us to probe the primordial power spectrum of adiabatic scalar perturbations with wavenumbers up to $k \simeq 0.2$~\mpc. We find no evidence for deviation from power-law fluctuations over two decades in scale. Matter fluctuations inferred from the primordial temperature power spectrum evolve over cosmic time and can be used to predict the matter power spectrum at late times; we illustrate the overlap of the matter power inferred from CMB measurements (which probe the power spectrum in the linear regime) with existing probes of galaxy clustering, cluster abundances and weak lensing constraints on the primordial power. This highlights the range of scales probed by current measurements of the matter power spectrum. \end{abstract}
\maketitle
\section{Introduction}
The cosmic microwave background (CMB) is light from the nascent universe, which probes early-universe physics. Measurements of the small-scale anisotropies of this radiation provide us with powerful constraints on many cosmological parameters, e.g., \citet{reichardt/etal:2009, sievers/etal:prep, komatsu/etal:prep, lueker/etal:2010, dunkley/etal:prep}.

In particular, the CMB constrains the power spectra of scalar and tensor perturbations, the relic observables associated
with a period of inflation in the early universe \citep{wang:etal/1999, tegmark/etal:2002Pk, bridle/etal:2003, mukherjee/wang:2003, easther/peiris:2006,kinney/kolb/melchiorri:2006,bridges/etal:2006, shafieloo/souradeep:2007,spergel/etal:2007wmap, verde/peiris:2008,reichardt/etal:2009, chantavat/etal:2009, bridges/etal:2009, peiris/etal:2010, vazquez/etal:2011}. The standard models of inflation predict a power spectrum of adiabatic  scalar perturbations close to scale-invariant. Such models are often described in terms of a spectral index $n_s$ and an amplitude of perturbations $\Delta^2_\mathcal{R}$ as $\mathcal{P}(k) = \Delta^2_{\mathcal{R}}\left( \frac{k}{k_0}\right)^{n_s -1},$ where $k_0$ is a pivot scale. A wide variety of models, however, predict features in the primordial spectrum of perturbations, which alter the fluctuations in the CMB \citep{amendola:etal/1994,kates:etal/1995,Atrio-Barandela:etal/1997,wang:etal/1999,einasto:etal/1999,kinney:2001,adams/etal:2001,matsumiya:etal/2002, blanchard/etal:2003, lasenby/doran:2003, hunt/etal:2007,  barnaby/etal:2009, achucaroo/etal:2010, nadathur/etal:2010,chantavat/etal:2009}, which can be constrained using reconstruction of the primordial power.

Primordial fluctuations evolve over cosmic time to form the large scale structures that we see today.  Therefore, a precision measurement of the power spectrum of these fluctuations, imprinted on the CMB, impacts all aspects of cosmology. Recent measurements of the CMB temperature and polarization spectra have put limits on the deviation from scale invariance including a variation in power-law with scale (e.g., a running of the spectral index, \citet{kosowsky/turner:1995});  in particular data from the Atacama Cosmology Telescope (ACT) \citep{das/etal:prep,dunkley/etal:prep} combined with \wmap\ satellite data \citep{larson/etal:prep} find no evidence for running of the spectral index with scale and disfavor a scale-invariant spectrum with $n_s = 1$ at $3\sigma$.

In this work we probe a possible deviation from power-law fluctuations by considering the general case where the power spectrum is parameterized as bandpowers within bins in wavenumber (or $k$) space. This `agnostic' approach allows for a general form of the primordial power spectrum without imposing any specific model of inflation on the power spectrum, and facilitates direct comparison with a wide range of models. Such tests of the primordial power have been considered by various groups \citep{wang:etal/1999, tegmark/etal:2002Pk, bridle/etal:2003, hannestad:2003, sealfon/etal:2005, spergel/etal:2007wmap, verde/peiris:2008, peiris/etal:2010, vazquez/etal:2011}. We revisit the calculation because of ACT high sensitivity over a broad range in angular scale.

This paper is based on data from 296 square degrees of  the \act\ 2008 survey in the southern sky, at a central frequency of 148 GHz. The resulting maps have an angular resolution of 1.4' and a noise level of between $25$ and $40~\mu$K per arcmin$^2$. A series of recent papers has described the analysis of the data and scientific results. The ACT experiment is described in \citet{swetz/etal:prep}, the beams and window functions are described in \cite{hincks/etal:2010}, while the calibration of the ACT data to \wmap\ is discussed in \citet{hajian/etal:prep}. The power spectra measured at \arone\ and \artwo\ are presented in \citet{das/etal:prep}, and the constraints on cosmological parameters are given in \citet{dunkley/etal:prep}. A high-significance catalog of clusters detected through their Sunyaev-Zel'dovich (SZ) signature is presented in \citet{marriage/etal:2010b}; the clusters are followed up with multi-wavelength observations described in \citet{menanteau/etal:2010}; the cosmological interpretation of these clusters is presented in \citet{sehgal/etal:2010b}. 
\vspace{2in}

 \section{Methodology}
\subsection{Angular Power spectrum\label{method}}
Following the work of \citet{wang:etal/1999,tegmark/etal:2002Pk, bridle/etal:2003, mukherjee/wang:2003} and \citet{spergel/etal:2007wmap}, we parameterize the primordial power spectrum $\mathcal{P}(k)$ using bandpowers in 20 bins, logarithmically spaced in mode $k$ from $k_1 = 0.001$ to $k_{20} = 0.35$~\mpc\ , with $k_{i+1} = 1.36k_i~\mathrm{for}~1<i<19$. To ensure the power spectrum is smooth within bins, we perform a cubic spline such that:
\newlength\mylen
\settowidth\mylen{$A_ib_{i} + Bb_{i+1} + \left((a^3-a)C_i +   (b^3-b)C_{i+1}\right)\frac{h_i^2}{6}$}

\begin{minipage}{2.9cm}
$\mathcal{P}(k) = \Delta^2_{\mathcal{R},0} \times \left\lbrace\vphantom{\rule{-1mm}{1.4cm}}\right.$ 
\end{minipage}
\begin{minipage}{\mylen}
$1  ~~~~~~~~~~~~~~~\text{ for $k < k_1$} $\\
\\
$ A_iP_{i} + B_iP_{i+1} + $\\
$ \left((A_i^3-A_i)C_i +   (B_i^3-B_i)C_{i+1}\right)\frac{h_i^2}{6}$\\
$~~~~~~~~~~~~~~~~~\text{ for  $k_{i}<k <  k_{i+1}$}$\\
\\
 $P_{20} ~~~~~~~~~~~~ \text{ for $k > k_{20}$} ~~~~~~~(1)$\\
\end{minipage}
\setcounter{equation}{1}
where the $P_i$ are the power spectrum amplitudes within bin $i,$ normalised so that $P_i = 1$ corresponds to scale invariance. The $\Delta^2_{\mathcal{R},0}$ is a normalized amplitude of scalar density fluctuations, which we take to be $2.36\times 10^{-9}$ \citep{larson/etal:prep}, and is the amplitude for a power-law spectrum around a pivot scale of $k_0 = 0.002$~\mpc. We do not vary the amplitude in our analysis as the power in the individual bands is degenerate with the overall amplitude; if a higher value was used, the estimated bandpowers would be lower by the corresponding amount, as we are measuring the total primordial power within a bin. The coefficients $C_i$ are the second derivatives of the input binned power spectrum data \citep{press/teukolsky/vetterling:NRC:2e}, $h_i = k_{i+1} - k_i$ is the width of the step and $A_i = (k_{i+1} - k)/h_i$ and $B_i = (k-k_{i})/h_i.$ We do not impose a `smoothness penalty' as discussed in \citet{verde/peiris:2008} and in \citet{peiris/etal:2010}. Adding more parameters to the parameter set makes it easier for the model to fit bumps and wiggles in the spectrum, hence as the number of bins increases, this parameterization will fit the noise in the data, particularly on large scales (small values of $k$). This in turn is expected to increase the goodness-of-fit of the model by approximately one per additional parameter. Hence, a model that fits the data significantly better than the standard $\Lambda$CDM power-law would yield an increase in the likelihood of more than one per additional parameter in the model. The logarithmic spacing in $k$ means that this is less of a problem at high multipoles, as many measurements are used to estimate the power in each band.

\begin{figure}[htbp!]
$\begin{array}{c}
\includegraphics[width=0.45\textwidth]{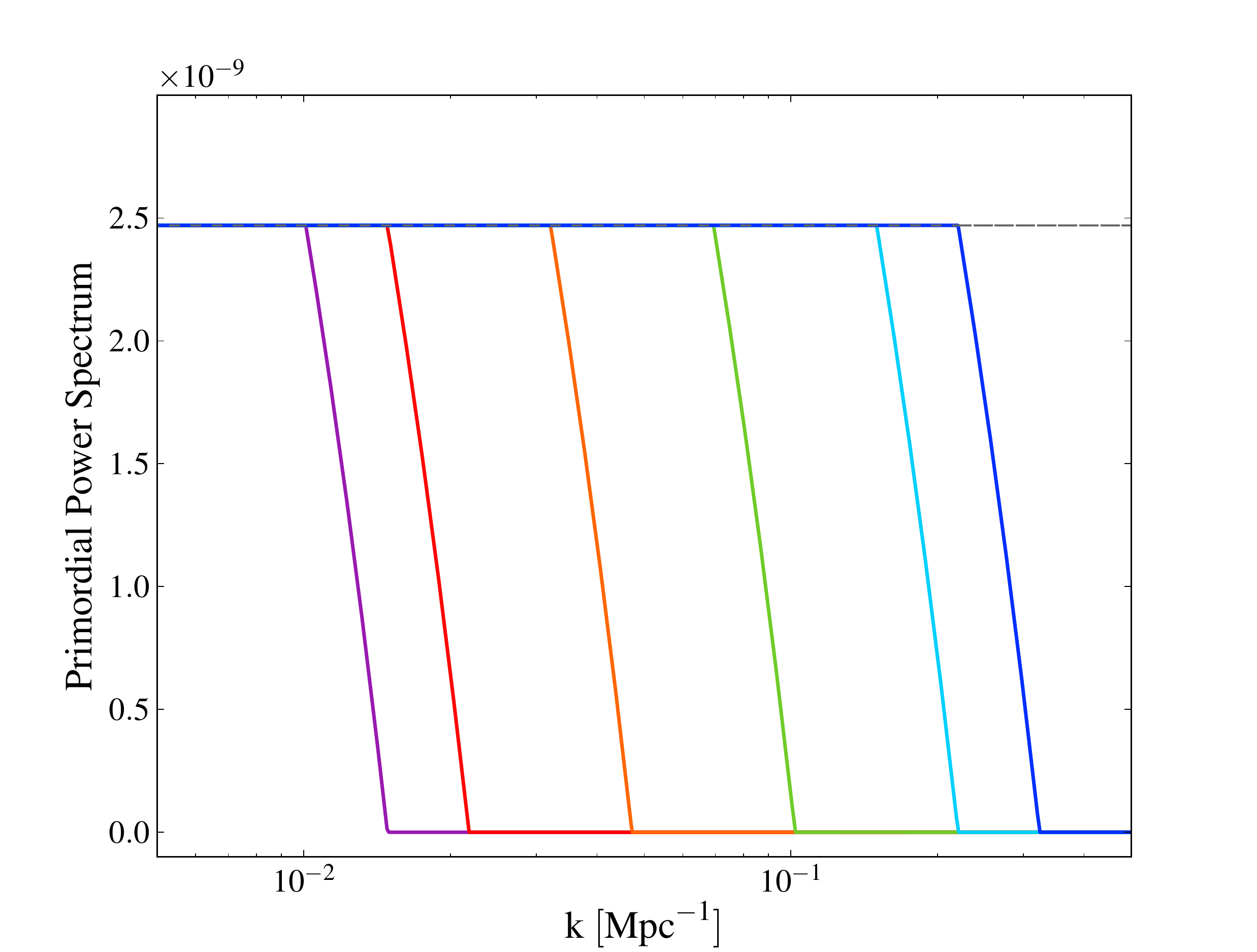} \\ [0.0cm]
\includegraphics[width=0.45\textwidth]{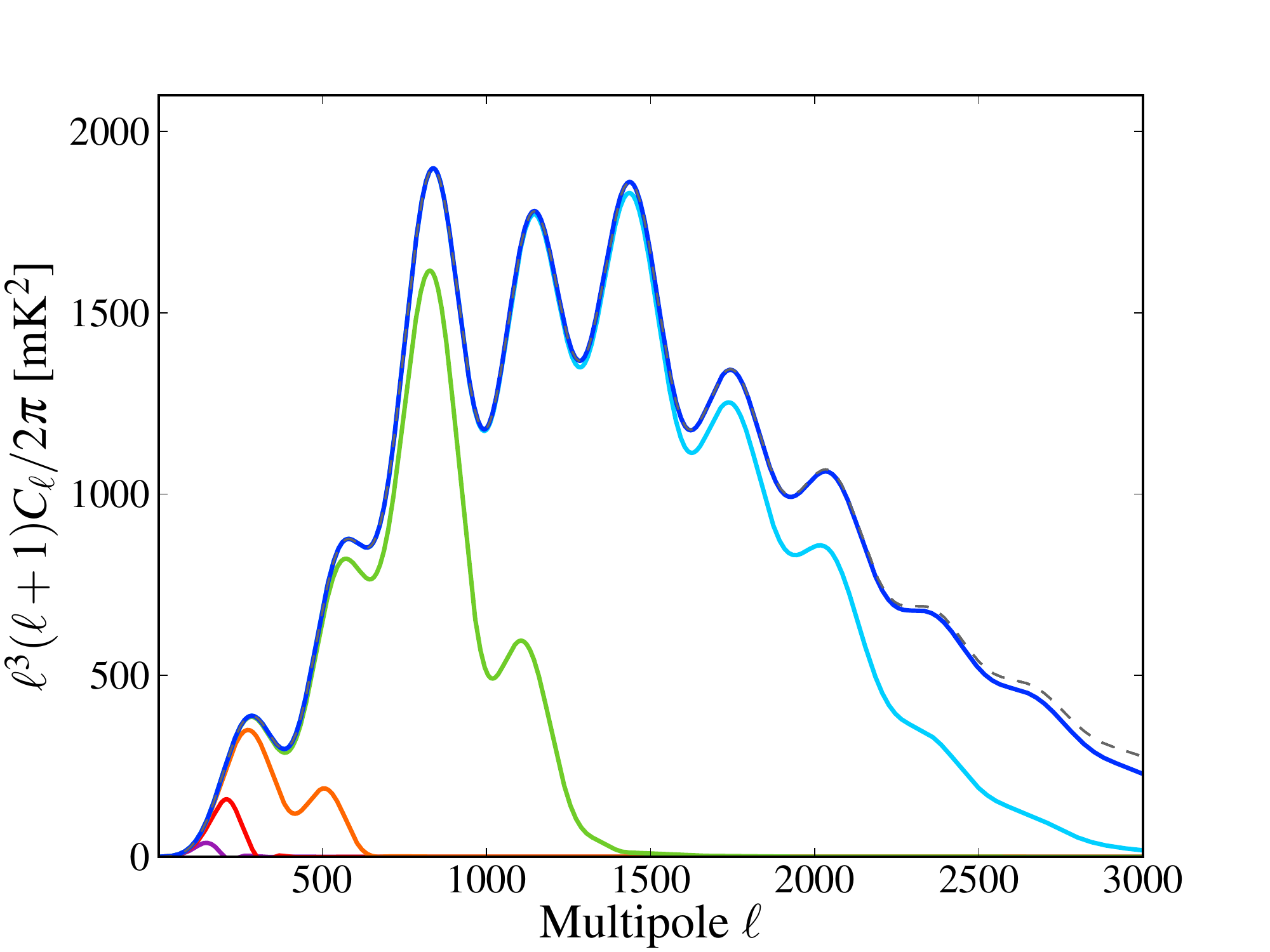} 
\end{array}$
\caption{Stepping up in power: we show schematically the angular power spectrum (lower panel) resulting from building up the primordial power spectrum $\mathcal{P}(k)$ in bins (top panel), from 
$k =  0.007$~\mpc\ (left-most curve in the top panel) to $k=0.22$~\mpc\ (right-most curve). The power in each case is normalized to a single amplitude before the step function, and set to zero afterwards, so that as more bins are added to the primordial spectrum, it tends towards a scale-invariant spectrum (shown as the dashed line). Correspondingly, the $C_\ell$ spectrum (plotted as $\ell^3(\ell+1)C^{TT}_\ell/2\pi$~mK$^2$ in the bottom panel) also tends to a spectrum characterized by $n_s=1$, also shown as the grey dashed curve. \label{fig:buildcmb}}
\end{figure}

The primordial power spectrum is related to the CMB power spectra through the radiation transfer functions $T_T(k)$, $T_{E}(k)$ and $T_B(k)$ (defined as in \citet{komatsu/spergel:2001}) as:

\begin{equation}
C_\ell^{\alpha \beta} \propto \int  k^2 dk \mathcal{P}(k)T_{\alpha}(k)T_{\beta}(k) \,
\label{eq:clpk}
\end{equation}

where $\alpha$ and $\beta$ index $T,E$ or $B$, corresponding to temperature or the two modes of polarization. The correspondence between multipole $\ell$ and mode $k$ (in~\mpc) is roughly $\ell \simeq kd$, where $d \simeq 14000$ Mpc is the comoving distance to the last scattering surface. 
Figure~\ref{fig:buildcmb} shows schematically how the primordial power spectrum translates to the temperature angular power spectrum. In each case, a single step function is used for the primordial power spectrum in Eq.~(\ref{eq:clpk}).

Previous analyses have only constrained the primordial power out to $k \lesssim 0.15$~\mpc\ \citep{bridle/etal:2003, spergel/etal:2007wmap, peiris/etal:2010}. The arcminute resolution of \act\ means that one can constrain the primordial power out to larger values of $k$ $(\simeq 0.19$~\mpc). The primary CMB power spectrum decreases exponentially due to Silk damping \citep{silk:1968} at multipoles greater than $\ell \simeq 2000$, while the power spectrum from diffuse emission of secondary sources begins to rise from $\ell \simeq 2000$. The \act\ measurement window between $1000<\ell<3000$ provides a new window with which to constrain any deviation from a standard power-law spectrum, as this is in the range of scales before the power from secondary sources dominates. We use the \arone\ measurements from the 2008 ACT Southern Survey, and include polarization and temperature measurements from the \wmap\ satellite with a relative normalization determined by \citet{hajian/etal:prep}. We use the ACT likelihood described in \citet{dunkley/etal:prep} and the \wmap\ likelihood found in \citet{larson/etal:prep}.

\subsection{Parameter estimation}

Our cosmological models are parameterized using:
\be
\Omega_c h^2, \Omega_b h^2, \theta_A, \tau, {\mathbf{P}},
\ee
where $\Omega_c h^2$ is the cold dark matter density; $\Omega_b h^2$ is the baryon density; $h$ is the dimensionless Hubble parameter such that $H_0 = 100h$ kms$^{-1}$\mpc; $\theta_A$ is the ratio of the sound horizon to the angular diameter distance at last scattering, and is a measure of the angular scale of the first acoustic peak in the CMB temperature fluctuations; $\tau$ is the optical depth at reionization, which we consider to be `instantaneous' (equivalent to assuming a redshift range of $\Delta z = 0.5$ for CMB fluctuations) and $\bold{P}  = \{P_i\}, ~~ i=1,..,20,$ is the vector of bandpowers  where $P_i = 1~\forall ~ i$ describes a scale-invariant power spectrum. We assume a flat universe in this analysis. In addition, we add three parameters, $A_{\mathrm{SZ}}, A_p, A_c,$ to model the secondary emission from the Sunyaev-Zel'dovich effect from clusters, Poisson-distributed and clustered point sources respectively, marginalizing over templates as described in \citet{dunkley/etal:prep} and \citet{fowler/etal:2010}. We impose positivity priors on the amplitudes of these secondary parameters. We modify the standard Boltzmann code CAMB\footnote{http://cosmologist.info/camb}  \citep{camb} to include a general form for the primordial power spectrum, and generate lensed theoretical CMB spectra to $\ell = 4000$, above which we set the spectra to zero for computational efficiency, as the signal is less than $5\%$ of the total power. 

The likelihood space is sampled using Markov chain Monte Carlo methods. The probability distribution is smooth, single-moded and close to Gaussian in most of the parameters. These properties make the 27-dimensional likelihood space less demanding to explore than an arbitrary space of this size: the number of models
in the Markov chain required for convergence scales approximately linearly with the number
of dimensions. Sampling of the parameter space is performed using CosmoMC\footnote{http://cosmologist.info/cosmomc} \citep{cosmomc}. The analysis is performed on chains of length $N=200000.$ We sample the chains  and test for convergence following the prescription in \citet{dunkley/etal:2005}, using an optimal covariance matrix determined from initial runs.

\renewcommand{\thefootnote}{\alph{footnote}}

\begin{deluxetable*}{cccc}  
\tablecolumns{4}
\tablecaption{Estimated power spectrum bands in units of $10^{-9}$ \label{pktable}}
\tablehead{   
\colhead{Wavenumber $k$ (\mpc)} &
  \colhead{Power spectrum band  \tablenotemark{a}\tablenotemark{b}} &
  \colhead{\wmap\ only binned $\mathcal{P}(k)$} &
  \colhead{\makecell{ACT+\wmap\ binned $P(k)$ }} 
}
\startdata
\\
0.0010& $P_1$ &$4.99^{+ 1.79}_{ - 1.77}$ & $5.07\pm 1.82 $\\
0.0014& $P_2$&$ < 3.22$ &$ < 3.49$ \\
0.0019&$P_3$& $< 3.04$ &$ < 3.03$\\
0.0025&$P_4$&$< 4.34$ &$<4.15$ \\
0.0034&$P_5$&$3.32\pm 0.99$ &$3.52 \pm 1.05$ \\
0.0047&$P_6$&$2.31^{+0.60}_{-0.58}$ &$2.29\pm 0.64$ \\
0.0064&$P_7$&$2.21\pm 0.33$ &$2.27\pm0.31$ \\
0.0087&$P_8$&$2.43\pm 0.19$ &$2.48\pm{0.20}$ \\
0.0118&$P_9$&$2.29\pm0.15$ &$2.35\pm 0.15$ \\
0.0160&$P_{10}$&$2.31\pm 0.13$ &$2.37 \pm 0.12$ \\
0.0218&$P_{11}$&$2.20\pm 0.11$ &$2.28\pm 0.11$ \\
0.0297&$P_{12}$&$2.38\pm 0.14$ &$2.40 \pm 0.13$ \\
0.0404&$P_{13}$&$2.28\pm 0.23$ &$2.39\pm 0.23$ \\
0.0550&$P_{14}$&$1.98 \pm 0.20$ &$2.14\pm0.14$ \\
0.0749&$P_{15}$&$2.37\pm 0.53$ &$2.41^{+0.20}_{-0.28}$ \\
0.1020&$P_{16}$&$< 4.01$ &$2.20^{+0.71}_{-0.80}$ \\
0.1388&$P_{17}$&$-$ &$2.19^{+0.79}_{-0.87}$ \\
0.1889&$P_{18}$&$-$ &$<2.37$ \\
0.2571&$P_{19}$&$-$ &$<2.40$ \\
0.3500&$P_{20}$&$ -$ &$-$ \\
\enddata
\tablenotetext{a}{For one-tailed distributions, the upper 95\% confidence limit is given, whereas the 68\% limits are shown for two-tailed distributions.}
\tablenotetext{b}{The primordial power spectrum is normalized by a fixed overall amplitude $ \Delta^2_{\mathcal{R},0} = 2.36\times 10^{-9}$ \citep{larson/etal:prep}. }\end{deluxetable*}

We impose limiting values on the power spectrum bands $0< P_i < 10$ for all $i$. To avoid exploring regions of parameter space inconsistent with current astronomical measurements, we impose a Gaussian prior on the Hubble parameter today of $H_0 = 74.2 \pm 3.6$ from \citet{riessHubble}.

\begin{figure*}[htbp!]
\centering
\includegraphics[width=0.8\textwidth]{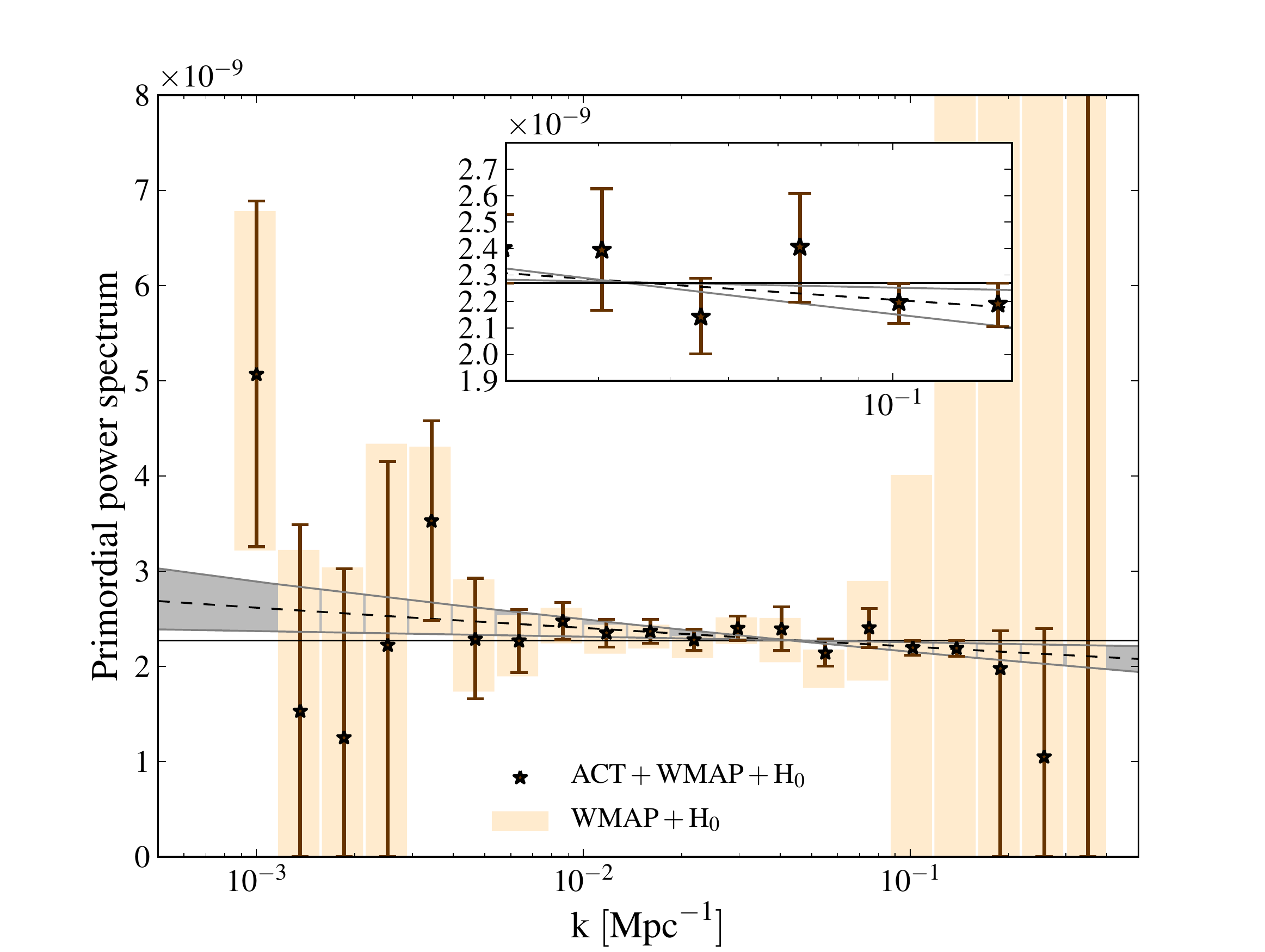}
\caption{Primordial power constraints: the constraints on the primordial power spectrum from the ACT data in addition to \wmap\ data compared to the \wmap\ constraints alone. In both cases, a prior on the Hubble parameter from \citet{riessHubble} was included. Where the marginalised distributions are one-tailed, the upper errorbars show the 95$\%$ confidence upper limits. On large scales the power spectrum is constrained by the \wmap\ data, while at smaller scales the ACT data yield tight constraints up to $k=0.19$~\mpc. The horizontal solid line shows a scale-invariant spectrum, while the dashed black line shows the best-fit $\Lambda$CDM power-law with $n_s = 0.963$ from \citet{dunkley/etal:prep}, with the spectra corresponding to the $2\sigma$ variation in spectral index indicated by solid band. The constraints are summarized in Table~\ref{pktable}. \label{fig:primkplot}}
\end{figure*}

\section{Results}
\subsection{Primordial Power \label{primk}}
Figure~\ref{fig:primkplot} shows the constraints on the primordial power spectrum from measurements of the cosmic microwave background. The shaded bands are the constraints on the power spectrum from \wmap\ measurements alone. Over this range of scales there is no indication of deviation from power-law fluctuations. As was shown in \citet{spergel/etal:2007wmap}, the lack of data at multipole moments larger than $\ell = 1000$ restricts any constraints on the primordial power spectrum at $k>0.1$~\mpc. In contrast, the combined ACT/\wmap\ constraints are significantly improved, particularly for the power at scales $0.1<k<0.19$~\mpc. 
The resulting power spectrum is still consistent with a power-law shape, with $n_s=0.963$ (the best-fit value from  \citet{dunkley/etal:prep}).  Despite the fact that we have added 18 extra degrees of freedom to the fit, a scale-invariant spectrum ($n_s=1$, shown by the horizontal line on Fig.~\ref{fig:primkplot}) is disfavored at $2\sigma$. In addition, we find no evidence for a significant feature in the small-scale power. The bands at $k > 0.19$~\mpc\ in the \act+\wmap\  case are largely unconstrained by the data. Including 218 GHz ACT data will improve the measurements of the primordial power, since it will relieve the degeneracies between the binned primordial power, the clustered IR source power, and the Poisson source power, all of which provide power at $\ell > 2000$.

\begin{figure*}[htbp!]
$\begin{array}{c c}
\includegraphics[width=0.5\textwidth]{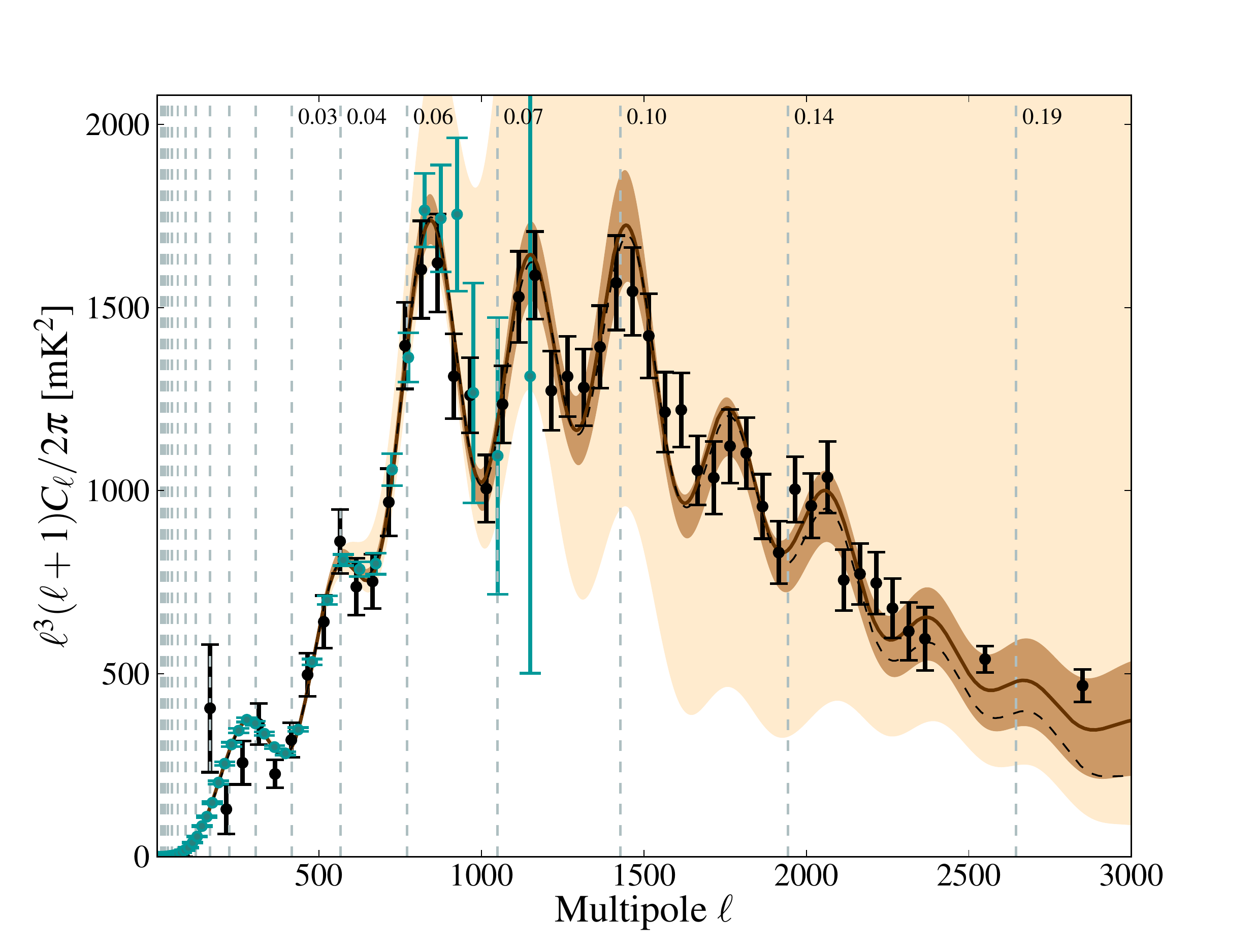} &
\includegraphics[width=0.5\textwidth]{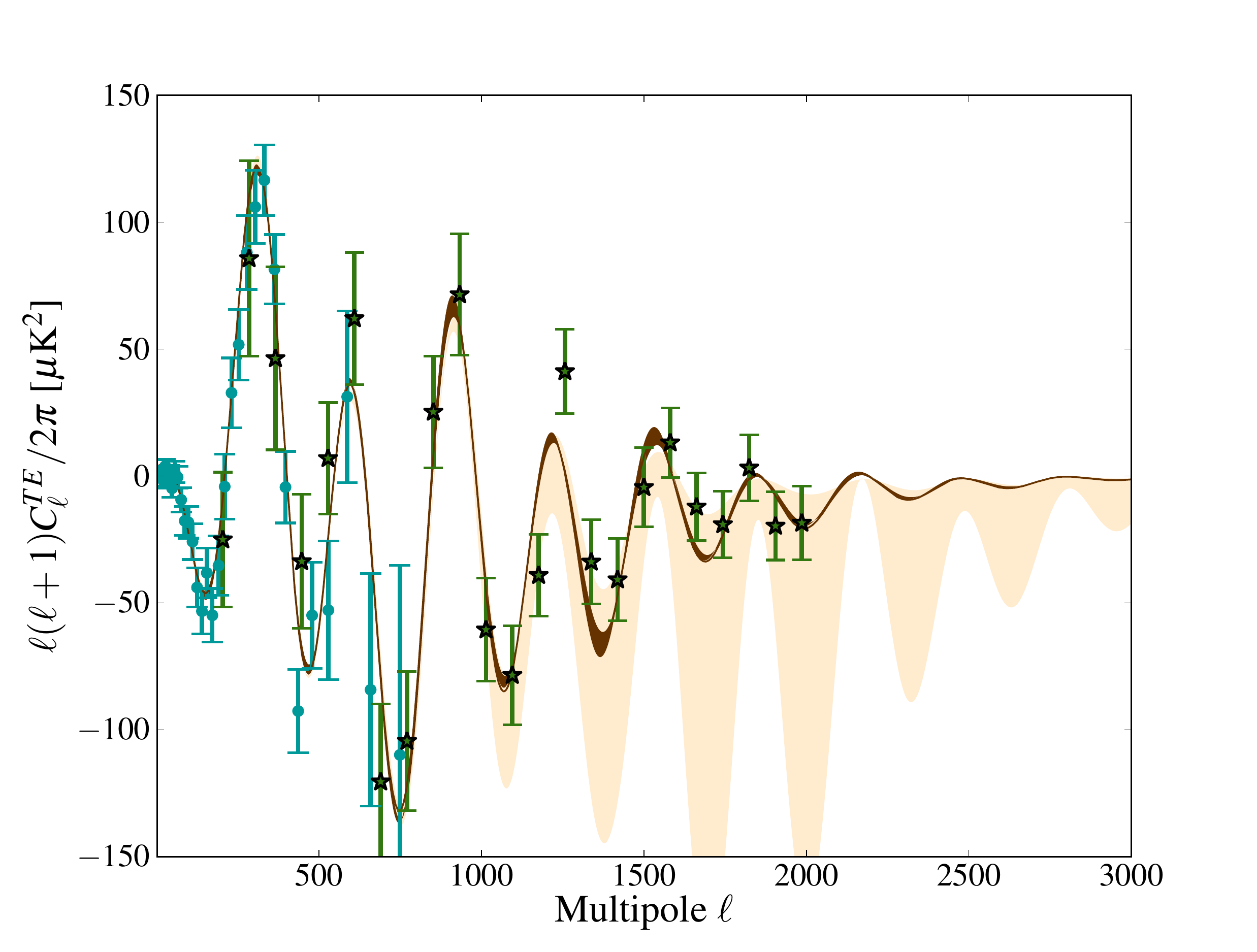} \\[0.0cm]
\end{array}$
\caption{Mapping primordial power to the angular power spectrum: the constraints on the primordial power spectrum from Figure ~\ref{fig:primkplot} translate into the angular power spectrum of the temperature CMB fluctuations, shown as $\ell^3(\ell+1)C^{TT}_\ell/2\pi$~mK${^2}$ (left panel) to highlight the higher order peaks. The dashed vertical lines show the multipoles corresponding to the wavenumbers under consideration, using $\ell=kd$; these wavenumbers as shown for the high$-k$ bands. The {dark (light)} band shows the $1\sigma$ region for the $C^{TT}_\ell$ spectra for the \act+\wmap\ (\wmap\ only) data. The best-fit curve using the combination of ACT and \wmap\ data is shown as the {{dark}} solid curve and the {{dashed black}} curve shows the best-fit power-law spectrum from \citet{dunkley/etal:prep}. The right panel shows the corresponding $C^{TE}_\ell$ power spectrum, plotted here as $\ell(\ell+1)C^{TE}_\ell/2\pi~\mu\mathrm{K}^{2},$ together with \wmap\ data and data from the QUaD experiment \citep{brown/etal:2010}.
\label{fig:clplot}}
\end{figure*}

The estimated primordial power spectrum values are summarized in Table~\ref{pktable}. The CMB angular power spectra corresponding to the allowed range in the primordial power spectrum (at $1\sigma$) are shown in Figure~\ref{fig:clplot} for \wmap-alone compared to \wmap\ and \act\ combined. The temperature-polarization cross spectra corresponding to these allowed models are also shown in Figure~\ref{fig:clplot}, indicating how little freedom remains in the small-scale $TE$ spectrum. This allowed range in $C_\ell^{TE}$ at multipoles $\ell > 1000$ will be probed by future CMB polarization experiments such as Planck \citep{ade/etal:2011}, ACTPol \citep{niemack/etal:2010} and SPTPol \citep{carlstrom/etal:2009}.

\begin{figure*}[htbp!]
$\begin{array}{c}
\hspace{-0.25in}
\includegraphics[trim = 0mm 50mm 0mm 50mm, clip, width=1.0\textwidth]{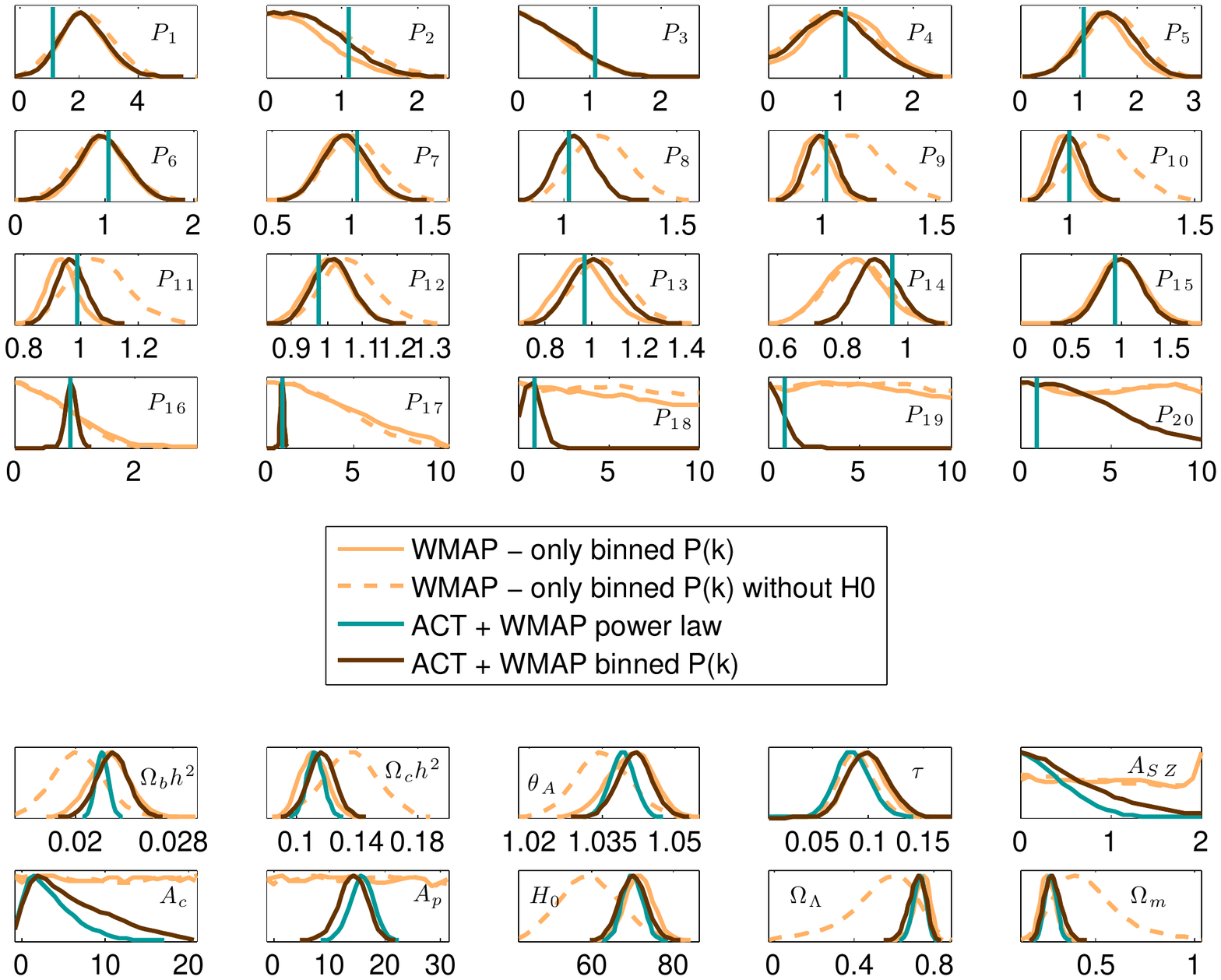}  \\[0.0cm]
\end{array}$
\caption{Parameter constraints: marginalized one dimensional distributions for the parameters determined from the ACT and \wmap\ data. The top 20 panels in the figure show the likelihoods for the power spectrum parameters directly determined using MCMC methods, while the lower 10 panels show the primary and secondary cosmological parameters and 3 derived quantities: the Hubble parameter $H_0$, the dark energy density $\Omega_\Lambda$, and the matter density $\Omega_m$. The light solid curves show the constraints on the parameters from ACT in combination with \wmap\ data for the $\Lambda$CDM case --- the vertical lines in the power spectrum panels show the values the power spectrum would take assuming the best-fit $n_s=0.963$ power-law from \citet{dunkley/etal:prep}. The parameter constraints for this power-law $\Lambda$CDM model is shown as the light curves. The solid {dark} lines show the distributions from ACT and \wmap\ data, assuming a prior on the Hubble constant. The best-fit value of the power-law spectral index obtained from fitting the well-constrained bands ($P_5 - P_{17}$) is $n_s = 0.965$. The dashed curves indicate the degeneracy between low values of $\theta_A$ and primordial power in modes around the position of the first peak.\label{fig:pkparam}}
\vspace{0.03in}
\end{figure*}

In this analysis, we use a prior on the Hubble constant.  Removing this prior reveals a degeneracy between the primordial power on scales $0.01 < k  < 0.02$~\mpc (bands $P_{8-11}$ in Figure~\ref{fig:pkparam}), and the set of parameters describing the contents and expansion rate of the universe.  Both affect the first acoustic peak. This degeneracy was previously noted in, e.g., \citet{blanchard/etal:2003, hunt/etal:2007,nadathur/etal:2010}, where a power spectrum model ``bump'' at $k=0.015$~\mpc\ was found to be consistent with observations in the context of a low $H_0$, and without any dark energy. Along this degeneracy, the primordial spectrum can be modified to move the position of the first peak to larger scales (relative to power-law), also increasing its relative amplitude. Since the first peak position is well measured by \wmap, this increase in angular scale is compensated by decreasing $\theta_A$. In a flat universe, this corresponds to a decrease in the Hubble constant and the cosmological constant. The matter density increases to maintain the first peak amplitude. The baryon density then decreases to maintain the relative peak heights. Imposing a prior on the Hubble constant has the effect of breaking this degeneracy. Alternatively, one could impose a prior on the baryon density from Big Bang Nucleosynthesis ($\Omega_bh^2 = 0.022 \pm 0.002$) which would disfavor the low-$H_0$ models, as indicated in the top left panel of the bottom rows in Figure~\ref{fig:pkparam}.

It is worth noting that the increase in the matter density along this degeneracy also increases the gravitational lensing deflection power, as a universe with a larger matter content exhibits stronger clustering at a given redshift. ACT maps have sufficient angular resolution to measure this deflection of the CMB (Das et al. 2011). Even without a strong prior on the Hubble constant, models with a bump in the primordial spectrum and $H_0<50$ (with $<10\%$ Dark Energy density) are disfavored at $>4\sigma$ from the lensing measurement alone, a result similar to that discussed in \citet{sherwin/etal:2011}. Although parameterized differently, the same argument applies to the primordial spectrum considered by \cite{hunt/etal:2007}, motivated by phase transitions during inflation, that eliminates dark energy but which is also disfavored at $\simeq 3\sigma$ by the lensing for a standard cold dark matter model.



\renewcommand{\thefootnote}{\alph{footnote}}

\begin{deluxetable*}{c c c c} 
\tablecolumns{4}
\tablecaption{Estimated model parameters and $68 \%$ confidence limits for the ACT 2008 Southern Survey data combined with \wmap\  \label{paramstable}}
\tablehead{   
\colhead{ } &
  \colhead{Parameter \tablenotemark{a}} &
  \colhead{ACT+\wmap\ Power-law\tablenotemark{b}} &
  \colhead{\makecell{ACT + \wmap\ binned $P(k)$\\ with $H_0$ prior} } 
}
\startdata
\\
Primary&$100 \Omega_bh^2$ &$2.222\pm 0.055$ &$2.307 \pm 0.124$ \\
&$\Omega_c h^2$ &$0.1125 \pm  0.0053$& $0.1166 \pm 0.0085$\\
&$\theta_A$ &$1.0394 \pm 0.0024$ & $1.0419 \pm 0.0034$\\
&$\tau$ & $0.086 \pm 0.014$& $0.100 \pm 0.017$\\
\\ \hline \\
Secondary &$A_p$ &$15.81 \pm 2.01$  & $14.19 \pm 2.45$ \\
&$A_c$ & $< 10.44$ &  $ < 17.08$ \\
& $A_{\mathrm{SZ}}$ & $ < 0. 92$& $<1.55$\\
\enddata
\tablenotetext{a}{For one-tailed distributions, the upper 95\% confidence limit is given, whereas the 68\% limits are shown for two-tailed distributions.}
\tablenotetext{b}{The power-law model for the primordial spectrum is $ \mathcal{P}(k) = \Delta^2_{\mathcal{R}}\left( \frac{k}{k_0}\right)^{n_s -1} $}
\label{tble:param}
\end{deluxetable*}

The estimated cosmological parameters are given in Table~\ref{tble:param} and the marginalized one-dimensional likelihoods are shown in Figure~\ref{fig:pkparam}.
While the binned $P(k)$ model adds 18 additional parameters to the parameter set, only 13 of those parameters are well constrained. All cosmological parameters in the binned power spectrum model are consistent with those derived using the concordance 6-parameter model with a power-law primordial spectrum. The addition of 13 new parameters which are substantially constrained by the data increases the likelihood of the model such that $-2\ln\mathcal{L} = 9.3.$ 
Using a simple model comparison criterion like the Akaike Information Criterion \citep[e.g.,][]{liddle:2004,takeuchi:2000}, the binned power spectrum model is disfavored over the standard concordance model. Further, we find a power-law slope fit to the 13 constrained bands in power spectrum space (bands labeled from $P_5$ to $P_{17}$ in Figure~\ref{fig:pkparam}) of $n_s = 0.965$, which is, as expected, consistent with the constraints on the spectral index from \citet{dunkley/etal:prep}.

\subsection{Reconstructed $P(k)$\label{latepk}}
\begin{figure*}[htbp!]
\begin{center}
$\begin{array}{c}
\includegraphics[width=0.8\textwidth]{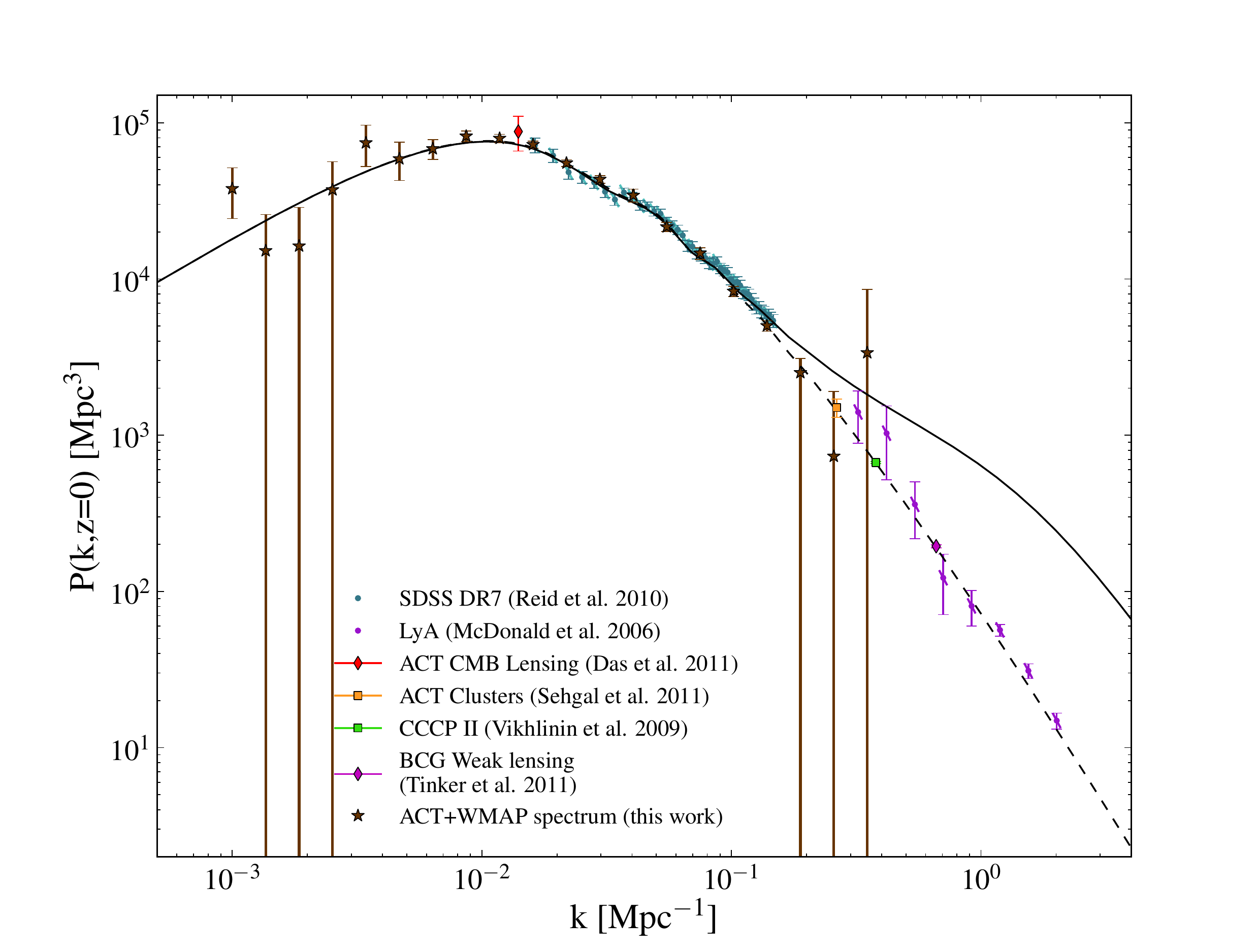}  \\ [0.0cm]
\includegraphics[width=0.8\textwidth]{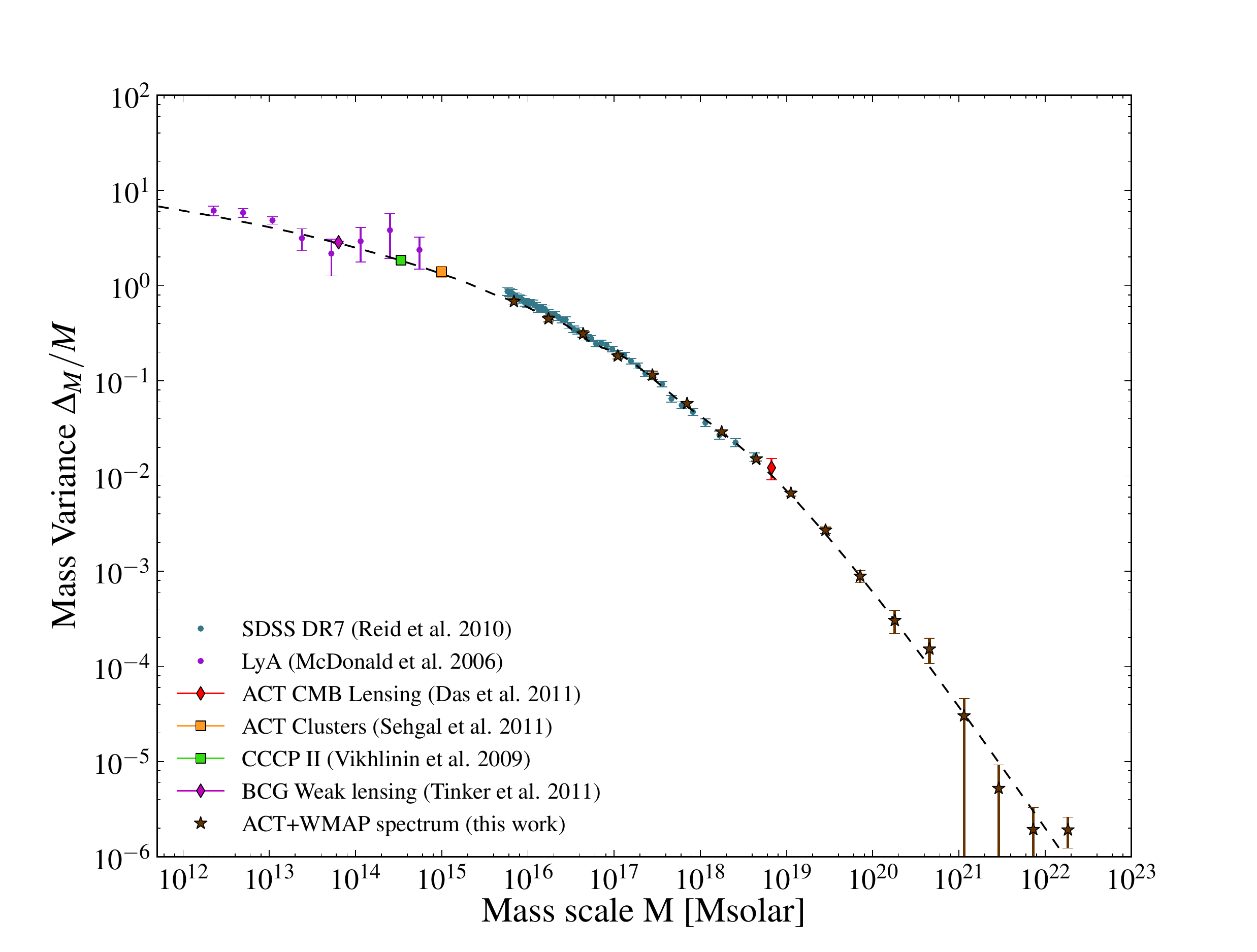}  
\end{array}$
\caption{The reconstructed matter power spectrum: the stars show the power spectrum from combining ACT and \wmap\ data (top panel). The solid and dashed lines show the nonlinear and linear power spectra respectively from the best-fit ACT $\Lambda$CDM model with spectral index of $n_s = 0.96$ computed using CAMB and HALOFIT \citep{smith/etal:2002}. The data points between $0.02< k < 0.19$~\mpc\ show the SDSS DR7 LRG sample, and have been deconvolved from their window functions, with a bias factor of 1.18 applied to the data. This has been rescaled from the \citet{reid/etal:2010} value of 1.3, as we are explicitly using the Hubble constant measurement from \citet{riess/etal:2011} to make a change of units from $h^{-1}$Mpc to Mpc. The constraints from CMB lensing \citep{dasetal:2011}, from cluster measurements from ACT \citep{sehgal/etal:2010b}, CCCP \citep{vikhlinin/etal:2009} and BCG halos \citep{tinker/etal:2011}, and the power spectrum constraints from measurements of the Lyman--$\alpha$ forest \citep{mcdonald/etal:2006} are indicated. The CCCP and BCG masses are converted to solar mass units by multiplying them by the best-fit value of the Hubble constant, $h=0.738$ from \citet{riess/etal:2011}. The bottom panel shows the same data plotted on axes where we relate the power spectrum to a mass variance, $\Delta_M/M,$ and illustrates how the range in wavenumber $k$ (measured in~\mpc) corresponds to range in mass scale of over 10 orders of magnitude. Note that large masses correspond to large scales and hence small values of $k$. This highlights the consistency of power spectrum measurements by an array of cosmological probes over a large range of scales. \label{fig:pkplot}}
\end{center}
\end{figure*}

The primordial power spectrum translates to the angular power spectrum of the CMB, but can in addition be mapped to the late-time matter power spectrum through the growth of perturbations: 
\begin{eqnarray}P(k,z=0) 
&=& 2\pi^2k\mathcal{P}(k)G^2(z)T^2(k)\, \end{eqnarray}

where $G(z)$ gives the growth of matter perturbations, $T(k)$ is the matter transfer function, and the $\mathcal{P}(k)$ are the fitted values as in Eq.~(1). This mapping enables the constraints on the power spectrum from the CMB to be related to power spectrum constraints from other probes at $z\simeq0$ \citep{tegmark/etal:2002Pk, birdetal:2010}. 
We illustrate the power spectrum constraints from the ACT and \wmap\ data in Figure~\ref{fig:pkplot}. We take $G(z)$ and $T(k)$ from a $\Lambda$CDM model, but neither varies significantly as the cosmological parameters are varied within their errors in the flat cosmology we consider in this work. The $P(k)$ constraints from the CMB alone overlap well with the power spectrum measurements from the SDSS DR7 LRG sample \citep{reid/etal:2010}, which have been deconvolved from their window functions. The \act\ data allow us to probe the power spectrum today at scales $0.001 < k < 0.19$~\mpc\ using only the CMB, improving on previous constraints using microwave data. In addition, the lensing deflection power spectrum also provides a constraint on the amplitude of matter fluctuations at a comoving wavenumber of $k\simeq 0.015$~\mpc\ at a redshift $z\simeq2$. The recent measurement of CMB lensing by ACT \citep{dasetal:2011} is shown as $P(k=0.015$~\mpc$) = 1.16\pm 0.29$~Mpc$^{3}$ on Figure~\ref{fig:pkplot}.  These two measurements are consistent with each other and come from two independent approaches: the lensing deflection power is a direct probe of the matter content at this scale (with only a minor projection from $z=2$ to $z=0$), while the primordial power is projected from the scales at the surface of last scattering at $z\simeq1040$ to the power spectrum today.

Finally, cluster measurements provide an additional measurement of the matter power spectrum on a characteristic scale $k_c$, corresponding to the mass of the cluster, $M_c = (4\pi/3)\rho_m (\pi/k_c)^3,$ where $\rho_m$ is the matter density of the universe today. We compute the amplitude of the power spectrum at the scale $k_c$ from reported $\sigma_8$ values as \vspace{-0.1in}\be P_c(k_c,z=0) = (\sigma_8/\sigma_{8,\Lambda \mathrm{CDM}})P(k_c,z=0)_{\Lambda \mathrm{CDM}}\,,\ee where $\sigma_{8,\Lambda \mathrm{CDM}} = 0.809$ is the concordance $\Lambda$CDM value \citep{larson/etal:prep}. We use the measurement of $\sigma_8 = 0.851 \pm 0.115$ from clusters detected by ACT, at a characteristic mass of $M = 10^{15} M_\odot$ \citep{sehgal/etal:2010b}, as well as measurements from the Chandra Cosmology Cluster Project (CCCP) \citep{vikhlinin/etal:2009}, measured from the 400 square degree ROSAT cluster survey \citep{burenin/etal:2006}. The quoted value of $\sigma_8 = 0.813\pm 0.013$ is given at a characteristic mass of $2.5 \times 10^{14}$ $h^{-1} M_\odot.$ In addition, we illustrate constraints from galaxy clustering calibrated with weak lensing mass estimates of brightest cluster galaxies (BCG) \citep{tinker/etal:2011}, quoted as $\sigma_8 = 0.826 \pm 0.02.$ In this case, we compute the characteristic mass $M_c$ (and hence $k_c$) from the inverse variance weighted average mass of the halos (from Table 2 in \citet{tinker/etal:2011}) as $M_c = 4.7 \times 10^{13}~h^{-1}M_\odot$. To remove the dependence on cosmology, the CCCP and BCG mass measurements are multiplied by a factor of $h^{-1}$ (where $h = 0.738$ is taken from the recent \citet{riess/etal:2011} result). The ACT cluster measurement, however, is already expressed in solar mass units, and hence this operation is not required. 

Power spectrum constraints from measurements of the Lyman--$\alpha$ forest are shown at the smallest scales probed. The slanted errorbars for the SDSS and Lyman--$\alpha$ data reflect the uncertainty in the power spectrum measurement from the Hubble constant uncertainty alone. Again, these data are normally plotted as a function of $h^{-1}$Mpc, hence we propagate the $1\sigma$ error on the Hubble parameter from \citet{riess/etal:2011} through to the plotted error region in both wavenumber and power spectrum. 

Transforming from units of power spectrum to mass variance $\Delta_M/M= \sqrt{P(k) k^3/(2\pi^2)}$ (indicated in the bottom panel of Figure~\ref{fig:pkplot}), allows one to visualize directly the relationship between mass scale and variance. While $\Delta_M/M \simeq 1$ for $10^{16} M_\odot$ galaxies, the variance decreases as the mass increases and we probe the largest scales, covering ten orders of magnitude in the range of masses of the corresponding probes.\

\section{Conclusions \label{conc}}
We constrained the primordial power spectrum as a function of scale in 20 bands using a combination of data from the 2008 Southern Survey of the Atacama Cosmology Telescope and \wmap\ data. We make no assumptions about the smoothness of the power spectrum, beyond a spline interpolation between power spectrum bands. The arcminute resolution of ACT constrains the power spectrum at scales $0.1< k < 0.19$~\mpc\ which had not yet been well constrained by microwave   background experiments. This allows us to test for deviations from scale invariance in a model-independent framework. We find no significant evidence for deviation from a power-law slope. When a power-law spectrum is fit to our well-constrained bands, our best-fit slope of $n_s=0.965$ is consistent with that determined
directly from a standard parameter space of $\Lambda$CDM models with a power-law spectrum,
using the same data. Mapping the primordial power to the late-time power spectrum using the fluctuations in the matter density, we obtain measurements of  the power spectrum today from the cosmic microwave background which are consistent with results from galaxy redshift surveys, but which also probe the power spectrum to much larger scales, $k \simeq 0.001$~\mpc, over mass ranges $10^{15} - 10^{22}~M_\odot$. Finally, the allowed range in the primordial power from the high-$\ell$ ACT temperature power spectrum measurements constrains the allowed range in the polarization-temperature cross spectrum, which will be probed with future polarization experiments.
\begin{acknowledgments}This work was supported by the U.S. National Science Foundation through awards AST-0408698 for the ACT project, and PHY-0355328, AST-0707731 and PIRE-0507768. Funding was also provided by Princeton University and the University of Pennsylvania, Rhodes Trust (RH), RCUK Fellowship (JD), ERC grant 259505 (JD), NASA grant NNX08AH30G (SD, AH and TM), NSERC  PGSD scholarship (ADH),
NSF AST-0546035 and AST-060697 (AK), NSF Physics Frontier Center grant PHY-0114422 (ES),
SLAC no.  DE-AC3-76SF0051 (NS), and the Berkeley Center for Cosmological Physics (SD)
Computations were performed on the GPC supercomputer at the SciNet HPC Consortium.
We thank Reed Plimpton, David Jacobson, Ye Zhou, Mike Cozza, Ryan Fisher, Paula Aguirre, Omelan Stryzak and the Astro-Norte group for assistance with the ACT observations. We also thank Jacques Lassalle and the ALMA team for assistance with observations. RH thanks Seshadri Nadathur for providing the best-fit power spectrum void models and Chris Gordon, David Marsh and Joe Zuntz for useful discussions. 
ACT operates in the Chajnantor Science Preserve in northern Chile under the auspices of the
Commission Nacional de Investigaci—n Cientifica y Tecnol—gica (CONICYT). Data acquisition electronics were developed with assistance from the Canada Foundation for Innovation.
\end{acknowledgments}
\bibliography{act,wmap_jo,wmap,pk}

\end{document}